\documentclass[paper,a4paper,11pt,twoside]{JHEP3}
\usepackage{psfrag}
\usepackage{epsfig}
\usepackage{amsmath}
\usepackage{axodraw}
\usepackage{cite}

\def\gev{\,\mathrm{GeV}}
\def\mev{\,\mathrm{MeV}}
\def\fm{\,\mathrm{fm}}
\def\SU{\mathrm{SU}}
\def\su#1#2{\SU(#1)_\mathrm{#2}}
\def\rpisq{\langle r_\pi^2\rangle}
\def\rpisqsu2sim{0.354(31)}  
\def\rpisqsimlong{0.382(42)}
\def\rpisqphys{0.418(31)} 
\def\rpisqphyslong{0.418(31)}
\def\lsixr{-0.0093(10)} 

\title{The pion's electromagnetic form factor at
	small momentum transfer in full lattice QCD}
\author{%
P.A.~Boyle$\,^a$, J.M.~Flynn$\,^b$, A.~J\"uttner$\,^c$,
 C.~Kelly$\,^a$, H.~Pedroso~de~Lima$\,^b$, C.M.~Maynard$\,^d$,
 C.T.~Sachrajda$\,^b$, J.M.~Zanotti$\,^a$ \\
$\,^a$SUPA, School of Physics, The University of Edinburgh,\\
\;\;Edinburgh, EH9 3JZ, UK\\
$\,^b$School of Physics and Astronomy, University of Southampton,\\
\;\;Southampton, SO17 1BJ, UK\\
$\,^c$Institut f\"ur Kernphysik, Johannes-Gutenberg Universit\"at Mainz,\\
\;\;D--55116 Mainz, Germany\\
$\,^d$EPCC, School of Physics, The University of Edinburgh,\\
\;\;Edinburgh, EH9 3JZ, UK}
\author{RBC and UKQCD Collaborations}

\preprint{%
Edinburgh 2008/18\\
MKPH--T--08--07\\
SHEP--08--15}

\abstract{%
  We compute the electromagnetic form factor of a ``pion" with mass
  $m_\pi=330\mev$ at low values of $Q^2\equiv -q^2$, where $q$ is the
  momentum transfer. The computations are performed in a lattice
  simulation using an ensemble of the RBC/UKQCD collaboration's gauge
  configurations with Domain Wall Fermions and the Iwasaki gauge
  action with an inverse lattice spacing of $1.73(3)\gev$. In order to
  be able to reach low momentum transfers we use partially twisted
  boundary conditions using the techniques we have developed and
  tested earlier. For the pion of mass $330\mev$ we find a charge
  radius given by $\rpisq_{330\,\textrm{MeV}}=\rpisqsu2sim\fm^2$ which,
  using NLO $\SU(2)$ chiral perturbation theory, extrapolates to a
  value of $\rpisq=\rpisqphys\fm^2$ for a physical pion, in agreement
  with the experimentally determined result. We confirm that there
  is a significant reduction in computational cost when
  using propagators computed from a single time-slice stochastic
  source compared to using those with a point source; for
  $m_\pi=330\mev$ and volume $(2.74\fm)^3$ we find the reduction is
  approximately a factor of $12$.}


\begin{document}

\section{Introduction}\label{sec:intro}

In this paper we compute the electromagnetic form factor of a ``pion"
with mass $m_\pi=330\mev$ at low values of $Q^2\equiv -q^2$, where $q$
is the momentum transfer. The computations are performed in a lattice
simulation using an ensemble of the RBC/UKQCD collaboration's gauge
configurations with Domain Wall Fermions and the Iwasaki gauge action
with an inverse lattice spacing of $1.73(3)\gev$ (see
section~\ref{subsec:parameters} for brief details of the simulation
and ref.~\cite{Allton:2008pn} for a full discussion\,\footnote{In ref.~\cite{Allton:2008pn} the pion mass corresponding to the bare quark mass used in the present study was found to be 331(6)\,MeV, where the error is dominated by the uncertainty in the lattice spacing. In the text we refer to this meson as having a mass of 330\,MeV, while in the analysis we treat the fluctuations in the mass using a jackknife procedure.}). The action has
good chiral and flavour symmetries and as demonstrated in
ref.~\cite{Allton:2008pn} a mass of $330\mev$ is well within the
regime where NLO $\SU(2)$ chiral perturbation theory holds for other
physical quantities, such as the meson masses, decay constants and the
kaon's bag parameter.

In order to be able to reach low momentum transfers we employ
partially twisted boundary conditions using the techniques developed
and tested in ref.~\cite{Boyle:2007wg}. Previous lattice computations
have used quarks satisfying periodic boundary conditions and therefore
obtained form factors at much larger values of $Q^2$ (see however, the
preliminary study with twisted boundary conditions presented by the
European Twisted Mass Collaboration (ETMC)~\cite{Simula:2007fa}). For the
pion with $m_\pi = 330\mev$ we find for the charge radius,
$\rpisq_{330\mev}=\rpisqsu2sim\fm^2$\,. We then use NLO chiral
perturbation theory to obtain the form factor and charge radius of a
physical pion, finding
\begin{equation}\label{eq:rpisqfinal}
\rpisq=\rpisqphys\fm^2\,,
\end{equation}
in agreement with the experimentally determined value.

The power of the technique is demonstrated in fig.~\ref{fig:fpipi} where the data points are obtained from our simulation. The dashed vertical line is the minimum value of $Q^2$ ($Q_{\textrm{min}}^2$) which is accessible with periodic boundary conditions. From the figure we see that the form factor can  be obtained at arbitrarily small values of $Q^2$ and also that the results obtained with twisted boundary conditions join smoothly onto those obtained by performing the Fourier sum in the conventional way (i.e. onto the data point on the dashed line). In this paper we focus on the pion's electromagnetic form factor, but we anticipate that the technique used here will also have important applications to the calculation of other flavour non-singlet form factors at arbitrary values of momentum transfer, such as those which appear in $K_{\ell 3}$ semileptonic decays~\cite{Boyle:2007wg}.

For this calculation we use propagators generated from a single
time-slice stochastic source in addition to standard point source
propagators. We compare the cost, obtaining similar errors for the
pion mass, the normalization constant of the vector current, $Z_V$, and
the pion's electromagnetic form factor at $Q^2_{\textrm{min}}$,
finding, for $m_\pi = 330\mev$ and volume $(2.74\fm)^3$, a gain of
approximately a factor of $12$ in favour of the noise source
propagators. A gain was also found in the preliminary study presented by the ETMC collaboration
in~\cite{Simula:2007fa} and in the recent
publication by the UKQCD collaboration~\cite{Boyle:2008rh}.

The plan for the remainder of this paper is as follows. In the next section we briefly review the use of partially twisted boundary conditions to compute the form factor at values of $Q^2$ which are inaccessible with periodic boundary conditions~\cite{Boyle:2007wg}. The details of the computation, the parameters of the simulation and the results for the form factor for the $330\mev$ pion are presented in section~\ref{sec:computation}. The use of noise source propagators to evaluate the three point correlation functions from which the form factor is obtained is briefly described in section~\ref{sct:Z2noise} and a comparison of the relative cost of using point source and noise source propagators to obtain results with the same statistical error is given in section~\ref{subsec:comparison}. The use of NLO chiral perturbation theory to obtain the form factor and charge radius for a physical pion from that with mass $330\mev$ is described and performed in section~\ref{sec:physical}. Finally, in section~\ref{sec:concs} we present our conclusions.

\section{Twisted boundary conditions and the form factor at small $Q^2$}\label{sec:twisted}

The electromagnetic form factor of the pion, $f^{\pi\pi}(q^2)$, is defined by the matrix element
\begin{equation}\label{eq:gen_ff}
\langle \pi^+(p^\prime)|V_\mu|\pi^+(p)\rangle=(p+p^\prime)_\mu\,f^{\pi\pi}(q^2),\qquad\textrm{where}\quad
		q^2=-Q^2=(p-p^\prime)^2
\end{equation}
and $V_\mu = \frac 23 \bar u \gamma_\mu u -
        \frac 13 \bar d \gamma_\mu d$ is the electromagnetic current.
In a finite volume with
periodic boundary conditions for the quark fields, the accessible pion momenta are given by
\begin{equation}
p=(E_n,\vec{p}_{\vec{n}})=(E_n,(2\pi/L)\,\vec{n})\quad\textrm{and}\quad p^\prime=(E_{n'},\vec{p}_{\vec{n}^\prime})=(E_{n'},(2\pi/L)\,\vec{n}^\prime)
\end{equation} where $\vec{n}$ and $\vec{n}^{\,\prime}$ are vectors of integers, $L$ is the spatial extent of the lattice and $E_n$ and $E_{n^\prime}$ are the corresponding energies ($E^2_n=m_\pi^2+(2\pi/L)^2\,|\vec{n}\,|^2$ and $E^2_{n^\prime}=m_\pi^2+(2\pi/L)^2\,|\vec{n}^{\,\prime}\,|^2$, where $m_\pi$ is the mass of the pion), so that $q^2$ can only take the corresponding discrete values. In particular the minimum non-zero value of $Q^2$ is given by $Q^2_{\rm min}=2m_\pi(\sqrt{m_\pi^2+(2\pi/L)^2}-m_\pi)$, which for the parameters of our simulation is about 0.15\,GeV$^2$. In this paper we study the form factor at small $Q^2$ (and in particular for $Q^2\ll 0.15\gev^2$), using the new technique proposed in \cite{Boyle:2007wg}
which allows one to carry out lattice computations at arbitrarily small
values of $Q^2$. We now briefly review this technique.
\begin{figure}
\begin{center}
\begin{picture}(120,60)(-60,-30)
\Line(-45,0)(-25,0)\ArrowLine(-37,0)(-36.5,0)
\Line(25,0)(45,0)\Oval(0,0)(12,25)(0)\ArrowLine(38,0)(38.5,0)
\GCirc(-25,0){3}{0.5}\GCirc(25,0){3}{0.5} \GCirc(0,12){3}{0.5}
\Text(-19,12)[b]{$q_1$}\Text(19,12)[b]{$q_2$}
\Text(0,-15)[t]{$q_3$}\Text(0,17)[b]{$V_\mu$}\Text(-49,0)[r]{$\pi(p)$}
\Text(49,0)[l]{$\pi(p')$}\ArrowLine(0.5,-12)(-0.5,-12)
\end{picture}
\end{center}
\caption{Sketch of the valence quark flow in the electromagnetic form
  factor of the pion. There is a similar contribution in which the
  current is on the antiquark line and the spectator is a
  quark.\label{fig:quarkflow}}
\end{figure}
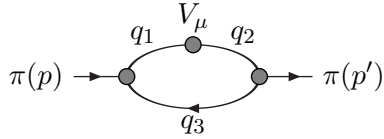

In order to reach small momentum transfers, we use partially twisted boundary conditions~\cite{Sachrajda:2004mi,Bedaque:2004ax},
combining gauge field configurations generated with sea quarks
obeying periodic boundary conditions with valence quarks with twisted
boundary conditions \cite{Bedaque:2004kc,deDivitiis:2004kq,Sachrajda:2004mi,
Bedaque:2004ax,Tiburzi:2005hg,Flynn:2005in,Guadagnoli:2005be,
Aarts:2006wt,Tiburzi:2006px,Bunton:2006va}. The valence quarks satisfy
\begin{equation}
q(x_k+L)=e^{i\theta_k}q(x_k),\qquad(k=1,2,3)\,,
\end{equation}
where $q$ represents one of the degenerate up or down quarks. We have demonstrated in section~2.3 of ref.~\cite{Boyle:2007wg} that it is possible to introduce twisted boundary conditions independently for the three valence quarks and antiquarks, i.e. $\vec\theta_1$ for $q_1$, $\vec\theta_2$ for $q_2$ and $\vec\theta_3$ for $q_3$ in fig.~\ref{fig:quarkflow}. In our study it will be sufficient to set $\vec\theta_3=0$ so that the spectator quark or antiquark satisfies periodic boundary conditions. By varying $\vec\theta_1$ and $\vec\theta_2$
we are able to tune the momenta of the initial and final pions continuously.

The dispersion relation for a meson with twisting angle $\vec\theta$ takes the form
\cite{deDivitiis:2004kq,Flynn:2005in},
\begin{equation}
 E_\pi =
 \sqrt{m_\pi^2 + \bigg(\vec p_{\vec{n}} + \frac{\vec \theta}L \,\bigg)^2},
\end{equation}
where $m_\pi$ is the pion mass and $\vec p_{\vec n}$ is the
meson momentum induced by Fourier summation. For the matrix element in (\ref{eq:gen_ff}) with the initial and
the final meson carrying momenta $\vec p=\vec{p}_{\vec{n}}+\vec\theta/L$ and $\vec p\,^\prime=\vec{p}_{\vec{n}\,^\prime}+\vec\theta\,^\prime/L$
respectively (where $\vec\theta=\vec\theta_1-\vec\theta_3$ and $\vec\theta\,^\prime=\vec\theta_2-\vec\theta_3$), the momentum transfer between the initial and the
final state meson is
\begin{equation}\label{eq:mom_transfer}
q^2=(p-p\,^\prime)^2=\Big(E_\pi(\vec p\,)-E_{\pi}
	(\vec p\,^\prime\,)\Big)^2
        -\left((\vec p_{\vec{n}}+\vec{\theta}/L)
          -(\vec p_{\vec{n}^\prime}+\vec{\theta}\,^\prime/L)
	\right)^2\ .
\end{equation}

The finite-volume corrections with partially twisted boundary conditions decrease exponentially with $L$ similarly to those with periodic boundary conditions~\cite{Sachrajda:2004mi}.

\section{The computation and results}\label{sec:computation}
In this section we present the details of the computation of the electromagnetic form factor of a pion with mass $m_\pi=330\mev$. In the first subsection we explain which correlation functions are computed in order to be able to extract the form factor. The parameters of the simulations are presented in subsection~\ref{subsec:parameters} and a brief introduction to the use of noise-source propagators is given in subsection~\ref{sct:Z2noise}. In section~\ref{subsec:f330} we present our results for the form factor. Finally in section~\ref{subsec:comparison} we compare the computational cost of computing correlation functions with point source and $\mathbb{Z}(2)$-wall source propagators.

\subsection{Correlation functions}
In order to determine the form factors we compute two- and
three-point correlation functions. The two-point function is
defined by
\begin{equation}
C_\pi(t,\vec p\,)=\sum_{\vec{x}}e^{i\vec{p}\cdot\vec{x}} \langle
\,O_\pi(t,\vec{x})\, O_\pi^\dagger(0,\vec{0})\,\rangle
    =\frac{
            |Z_\pi|^2}{2E_\pi(\vec p\,)}
            \left(e^{-E_\pi(\vec p\,) t}+ e^{-E_\pi(\vec p\,)(T-t)}\right)\, ,
\label{eq:twopt}\end{equation}
where $O_\pi= \bar d \gamma_5 u$ is a local pseudoscalar interpolating operator for the pion. We have assumed that $t$ and $T-t$ (where $T$ is
the temporal extent of the lattice) are sufficiently large for the
correlation function to be dominated by the lightest state (i.e. the
pion). The constant $Z_\pi$ is given by
$Z_\pi=\langle\,\pi\,|\,O_\pi^\dagger(0,\vec{0})\,|\,0\,\rangle$\,.
The three-point function is defined by
\begin{eqnarray}
C_{\pi\pi}(t,t_{f},\vec p,\vec p\,^\prime) &=&
Z_V\,\sum_{\vec{x}_f,\vec{x}} e^{i\vec{p}\,^\prime\cdot(\vec{x}_f-\vec{x})}
e^{i\vec{p}\cdot\vec{x}} \langle\, O_\pi(t_f,\vec x_f)\,
V_4(t,\vec{x})\,O_\pi^\dagger(0,\vec 0)\,\rangle\nonumber\\
    &=&
        \frac{Z_V\,|Z_\pi|^2}{4E_\pi(\vec p\,) E_{\pi}(\vec p\,^\prime)}\,
	\langle\,\pi(\vec{p}\,^\prime)\,|\,V_4(0)\,|\,
        \pi(\vec{p}\,)\,\rangle\,\nonumber\\[3mm]
&&\hspace{-1in}
        \times\left\{\theta(t_f-t)\,e^{-E_\pi(\vec{p}\,)\,t-
	E_{\pi}(\vec p\,^\prime)(t_f-t)}\ -\
	\theta(t-t_f)\,e^{-E_\pi(\vec p\,)(T-t)-
	E_{\pi}(\vec p\,^\prime)(t-t_f)}\right\}\,,\label{eq:threept}
\end{eqnarray}
where $V_4$ is the time component of the bare electromagnetic current
and where, without loss of generality, we have placed the source at
the origin. Again we assume that all the time intervals in~(\ref{eq:threept}) are sufficiently
large for the lightest hadrons to give the dominant contribution. As explained in the following paragraph, $Z_V$ is the normalization factor by which the bare lattice current needs to be
multiplied in order to obtain the physical current.

The normalization factor $Z_V$ can readily be obtained as follows. For illustration we take $0< t< t_f <T/2$, in which case $Z_V$ is defined by
\begin{equation}\label{eq:zv}
Z_V = \frac{\tilde C_\pi(t_f,\vec 0)}
    {C^B_{\pi\pi}(t,t_f,\vec{0},\vec{0}\,) }\,.
\end{equation}
In the numerator we use the function
$\tilde C_\pi(t,\vec p)=
	C_\pi(t,\vec p)-
	\frac{|Z_\pi|\,^2}{2E_\pi(\vec p)}\,e^{-E_\pi(\vec p) (T-t)}$
where $Z_\pi$ and $E_{\pi}(\vec p)=\sqrt{m_\pi^2+\vec p{\,^2}}$
are determined from fits to
$C_\pi(t,\vec 0\,)$.  For $t_f<T/2$ this proves to be an effective and numerically stable procedure for the subtraction of the contribution from the backward propagating meson to
$C_\pi(t_f,\vec 0\,)$ in the numerator of (\ref{eq:zv}). (For $t_f=T/2$ it is natural instead to use
$\tilde C_\pi(t,\vec p)=\frac 12 C_\pi(t,\vec p)$ in (\ref{eq:zv}).) The
superscript $B$ in the denominator indicates that we take the bare (unrenormalized)
current in the three-point function.

In the following subsection we introduce the three datasets which we
use for our analysis. For data set A we do not use twisted boundary
conditions, setting $\vec p\,'=0$ and determining the pion form factor
from the ratio of correlation functions
\begin{equation}\label{eqn:R_A}
 {2m_\pi}\frac{C_{\pi\pi}(t,t_f,\vec p,\vec 0)}{C_{\pi\pi}(t,t_f,\vec 0,\vec 0)}
        \frac{\tilde C_\pi(t,\vec 0)}{\tilde C_\pi(t,\vec p)}
	\longrightarrow
        f^{\pi\pi}(q^2)\left({E_\pi(\vec p)+m_\pi}\right)\,.
\end{equation}
For data sets B and C we use
\begin{equation}\label{eqn:R_1}
 2\sqrt{E_\pi(\vec{p}\,) E_\pi(\vec{p}\,^\prime)}\
 \sqrt{\frac
 {C_{\pi\pi}(t,t_f,\vec p,\vec{p}^{\,\prime})\,
	C_{\pi\pi}(t,t_f,\vec p^{\,\prime},\vec{p}\,)}
 {\tilde C_{\pi}(t_f,\vec p\,)\,\tilde C_{\pi}(t_f,\vec{p}^{\,\prime})}}\longrightarrow
	f^{\pi\pi}(q^2)(E_\pi(\vec p\,)+E_\pi(\vec p\,^\prime))\,,
\end{equation}
(called ratio $R_1$ in \cite{Boyle:2007wg}). Both ratios
approach a constant for sufficiently large time intervals.

\subsection{Parameters of the simulation}\label{subsec:parameters}

The computations described in this paper were performed using the
ensemble with light quark mass $am_u=am_d=0.005$ and strange quark mass $am_s=0.04$ from the set of
$N_f=2+1$ flavour Domain Wall
Fermion~\cite{Kaplan:1992bt,Shamir:1993zy,Furman:1994ky}
configurations with $(L/a)^3\times T/a\times L_s=24^3\times 64\times
16$ which were jointly generated by the UKQCD/RBC collaborations using
the QCDOC computer~\cite{qcdoc1,qcdoc2,qcdoc3,qcdoc4}.
The bulk of the
correlation functions were evaluated on the UK Research Councils'
HECToR Cray XT4 computer, with the set completed using a University of
Edinburgh BlueGene/L system as well as QCDOC. A detailed study of the
light-hadron spectrum and other hadronic quantities using these
configurations has recently been reported in
ref.\,\cite{Allton:2008pn}. The gauge configurations were generated
with the Iwasaki gauge action~\cite{Iwasaki:1985we,Iwasaki:1984cj} at
$\beta=2.13$ corresponding to an inverse lattice spacing of
$a^{-1}=1.729(28)\gev$. The resulting pion mass is $m_\pi \approx
330\mev$. In our numerical evaluations we use the masses
measured directly on our data sets, which can be determined from the
entries in tab.~\ref{tab:pipi_kinematics} and are fully
consistent with the value reported in~\cite{Allton:2008pn}. We use the
jackknife technique to estimate the statistical errors.

In the following we distinguish three sets of correlation functions as
specified in tab.~\ref{tab:sets}. Set A was generated with point
sources and sinks. We started the measurements for three different
source positions on trajectories $900$, $905$ and $910$, respectively,
measuring on every 40th trajectory in each case and averaging three
consecutive measurements over the sources into one bin. The initial
pion carries momentum $|\vec p\,|=0$, $\frac{2\pi}{L}$ or
$\sqrt{2}\frac{2\pi}{L}$ and the final pion is at rest. For this
dataset we do not use twisted boundary conditions at all.

For data sets B and C we used a $\mathbb{Z}(2)\times \mathbb{Z}(2)$
noise wall source as explained in section~\ref{sct:Z2noise}
and a point sink. For data set B we started the measurement chains for
the eight source positions specified in tab.~\ref{tab:sets} on
trajectories $900, 905, 910,\dots,935$. Data set C is a subset of set
B which starts with four source positions on trajectories
$900,910,920$ and $930$, respectively. In each case we measured on
every 40th trajectory and averaged the correlation functions over the
chains into bins of eight and four measurements, respectively. The
correlation functions obtained using sets B and C were computed with
$\vec p_{\vec{n}}=\vec{p}_{\vec{n}\,^\prime}=0$ and the
momenta of the initial and/or final pions were induced by twisting one
of the pions' valence quarks. For each measurement we applied the full
twist along one of the spatial directions. We changed this direction
frequently as the measurements proceeded in order to reduce the
correlations. In the cases in which both the initial and the final
pion carried a twist, $\vec\theta$ and $\vec\theta\,^\prime$ were chosen
to be anti-parallel.

Based on a preliminary study of a subset of data set A, we determined
the pion mass $am_\pi$ to have the central value $0.1907$ and this
guided us to choose twisting angles so as to obtain a suitable range
of momentum transfers. (After a detailed analysis by the RBC/UKQCD
collaboration on their data set of choice, called the FPQ data set in
\cite{Allton:2008pn}, the mass was quoted as $0.1915(8)$.) For such a
mass, as mentioned in section~\ref{sec:twisted}, the minimum value of
$Q^2$ which can be reached without using twisted boundary conditions
is $(aQ)^2_{\rm min}\approx0.051$ ($Q^2_{\rm min}\approx
0.152\gev^2$). In order to reach smaller values of $Q^2$ we introduce
three twisting angles $2.6832$, $2.1285$ and $1.6$, and in
tab.~\ref{tab:pipi_kinematics} we summarize the corresponding
kinematics.
\begin{table}
 \begin{center}
 \begin{tabular}{lcccccc}
 \hline\hline\\[-4mm]
 set&trajectories on $t_\mathrm{src}\!=\!0$& $\Delta$&$N_\mathrm{meas}$&
 $t_\mathrm{src}$\\[1mm]
 \hline\\[-4mm]
 A&900\,--\,4460&20&537&0,16,32\\
 B&1000\,--\,6840&40&1176&0, 54, 20, 14, 56, 26, 44, 34\\
 C&1000\,--\,6440&40&548&0, 20, 56, 44\\[1mm]
 \hline\hline
 \end{tabular}
 \end{center}
 \caption{Details of measurements A, B and C. The quoted range of
   trajectories is the one for $t_\mathrm{src}=0$ and $\Delta$ is the
   separation in units of trajectories between subsequent measurements
   for each source position $t_\mathrm{src}$.}\label{tab:sets}
\end{table}

 \begin{table}
 \begin{center}
 \begin{tabular}{cccccccc}
 \hline\hline\\[-4mm]
data set & $|\vec p\,|L$ & $|\vec p\,'|L$ & $aE_\pi(\vec p)$
 & $aE_{\pi}(\vec p\,')$ & $(aQ)^2$ & $Q^2\,(\gev^2)$ & $f^{\pi\pi}(q^2)$\\[1mm]
 \hline\\[-4mm]
B &          $0$ &           $0$    &    $0.1910(4)$  &   $0.1910(4)$  &   $0$        &  $0$      &      $1$            \\
B &          $0$ &           $1.6$  &    $0.1910(4)$  &   $0.2023(4)$  &   $0.004$    &  $0.013$  &      $0.9804(15)$          \\
B &          $0$ &         $2.1285$ &    $0.1910(4)$  &   $0.2106(4)$  &   $0.007$    &  $0.022$  &      $0.9660(24)$          \\
B &          $0$ &         $2.6832$ &    $0.1910(4)$  &   $0.2213(4)$  &   $0.012$    &  $0.035$  &      $0.9477(36)$          \\
C &          $1.6$ &        $1.6$   &    $0.2023(6)$  &   $0.2023(6)$  &   $0.018$    &  $0.053$  &      $0.9189(75)$          \\
C &          $2.1285$ &     $1.6$   &    $0.2106(5)$  &   $0.2023(6)$  &   $0.024$    &  $0.072$  &      $0.8943(88)$          \\
C &          $2.1285$ &    $2.1285$ &    $0.2106(5)$  &   $0.2106(5)$  &   $0.031$    &  $0.094$  &      $0.867(10)$          \\
C &          $2.6832$ &     $1.6$   &    $0.2213(5)$  &   $0.2023(6)$  &   $0.031$    &  $0.094$  &      $0.864(11)$          \\
C &          $2.6832$ &    $2.1285$ &    $0.2213(5)$  &   $0.2106(5)$  &   $0.040$    &  $0.120$  &      $0.838(12)$          \\
C &          $2.6832$ &    $2.6832$ &    $0.2213(5)$  &   $0.2213(5)$  &   $0.050$    &  $0.150$  &      $0.802(15)$          \\
\hline\\[-5mm]
A&0     &$0$            &0.1912(7)&0.1912(7)&0    &0    &1\\
A&$2\pi$&$0$            &0.3242(4)&0.1912(7)&0.051&0.152&0.809(14)\\
A&$\sqrt2\,2\pi$&0       &0.4167(3)&0.1912(7)&0.086&0.258&0.711(26)\\
 \hline\hline
 \end{tabular}
 \end{center}
\caption{Table of accessible values of $Q^2=-q^2$ for the matrix element
$\langle \pi(p^\prime)|V|\pi(p)\rangle$ together with the values of
$f^{\pi\pi}(q^2)$\,. For data set B and C we also determined the correlation
functions with momenta $|\vec p\,|L$ and $|\vec p\,{'}|L$ interchanged.}
\label{tab:pipi_kinematics}
\end{table}

\subsection{Three point functions from noise source propagators}
\label{sct:Z2noise}

Lattice quark propagators are calculated by inverting the Dirac matrix
$\mathcal{D}$ upon a matrix valued source $\eta$,
\begin{equation}
S_{A,C}(t,\vec{x};t_i)\equiv
 \sum_{\vec{y}}\sum_B \mathcal{D}^{-1}_{A,B}(t,\vec{x};t_i,\vec{y})
 \eta_{B,C}(\vec{y})\,,
\end{equation}
where $A,B,C$ are spin-colour indices.

The hadronic form factor calculation is traditionally performed using
point source propagators~\cite{Boyle:2007wg}, for which the Dirac
matrix is inverted from a single site with $\eta_{B,C}(\vec{y}) =
\delta_{\vec{y},\vec{0}}\;\delta_{B,C}$. However it has been
shown~\cite{Foster:1998vw,McNeile:2006bz,Boyle:2008rh} that the
use of stochastic sources allows for the calculation of meson
propagators at a substantially reduced cost.

Following~\cite{Dong:1993pk,Foster:1998vw,McNeile:2006bz} we use
source matrices with random elements from the set $\mathbb{Z}(2)$ for
both real and imaginary components on a \textit{single} source
spin-colour index ($0$), for all sites $\vec{y}$ on the source
timeslice: $\eta_{B,0}(\vec{y},t_i)\in \mathbb{Z}(2)\otimes
\mathbb{Z}(2)$. With sources of this form, the solution $S(\vec{x},
t;t_i)$ requires only a single inversion rather than the 12 required
for the point solution.

A set $\{\eta^{j}|j=1,\ldots, N\}$ of these sources has the property that in the limit $N\rightarrow \infty$
\begin{equation}
\frac1N
\sum_{j=0}^N \eta^j_{A,0}(\vec{x},t_i)\eta^{\dagger\;j}_{0,B}(\vec{y},t_i) \rightarrow \delta_{\vec{x},\vec{y}}\;\delta_{A,B}
\label{eqn:stochcancellation}
 \end{equation}
such that the pseudoscalar two-point correlator at zero momentum tends to the spatial average of the point source solution \cite{Foster:1998vw,McNeile:2006bz}
\begin{equation}
\begin{aligned}
C_{\pi}(t,\vec 0) &= \sum\limits_{j=0}^N \sum\limits_{\vec{x}}\mathrm{tr}\left\{\; \gamma^5 S^j(\vec{x},t; t_i) \gamma^5 \left(\gamma^5 S^j(\vec{x}, t;t_i)\gamma^5\right)^\dagger\right\}\\
   &\rightarrow \sum\limits_{\vec{x},\vec{y}}\mathrm{tr}\left\{\; \gamma^5 \mathcal{D}^{-1}(\vec{x},t; \vec{y}, t_i) \gamma^5 \left(\gamma^5 \mathcal{D}^{-1}(\vec{x}, t;\vec{y},t_i)\gamma^5\right)^\dagger\right\}\,.
\end{aligned}
\label{eqn:pscorrelator}
\end{equation}
Although this explicitly projects to zero momentum at source, twisted boundary conditions can be used to induce a non-vanishing meson momentum.

The properties of equation~(\ref{eqn:stochcancellation}) are retained on average when the sources $\eta^i$ reside on different \textit{configurations} such that the stochastic sum can be included in the ensemble average. Therefore we require only a single stochastic source per configuration, giving an overall factor of 12 cost reduction over the traditional method.

This technique can be extended simply to three-point correlators using standard sequential source methods
\begin{equation}
S^{\prime}(t_i;t_f,\vec {p}_f;t,\vec{x})
    =\sum_{\vec{x}_f}\gamma_5\left(\mathcal{D}^{-1}(t,\vec {x};t_f,\vec{x}_f)
        \gamma^5 S(t_f,\vec{x}_f;t_i)\,
        e^{-i\vec{p}_f\cdot\vec{x}_f}\right)^\dagger \gamma_5\,,
\end{equation}
the solution of which is again non-zero only on a single source
spin-colour index, thus requiring only one extra inversion. The
stochastic cancellation with the other source occurs at the source
timeslice $t_i$ as in (\ref{eqn:pscorrelator}).

\subsection{Electromagnetic form factor of a pion with
  \boldmath{$m_\pi=330\mev$}}
\label{subsec:f330}

\begin{figure}
\psfrag{xlabel}[c][b][1][0]{$Q^2[\gev^2]$}
\psfrag{ylabel}[c][t][1][0]{$f^{\pi\pi}(q^2)$}
\psfrag{set A}[c][c][1][0]{\small set A}
\psfrag{set B}[c][c][1][0]{\small set B}
\psfrag{set C}[c][c][1][0]{\small set C}
\psfrag{pole fit}[c][c][1][0]{\small pole fit}
\psfrag{QCDSF/UKQCD       }[l][lc][1][0]{\small \!\!QCDSF/UKQCD}
\begin{center}
\epsfig{scale=.47,angle=-90,file=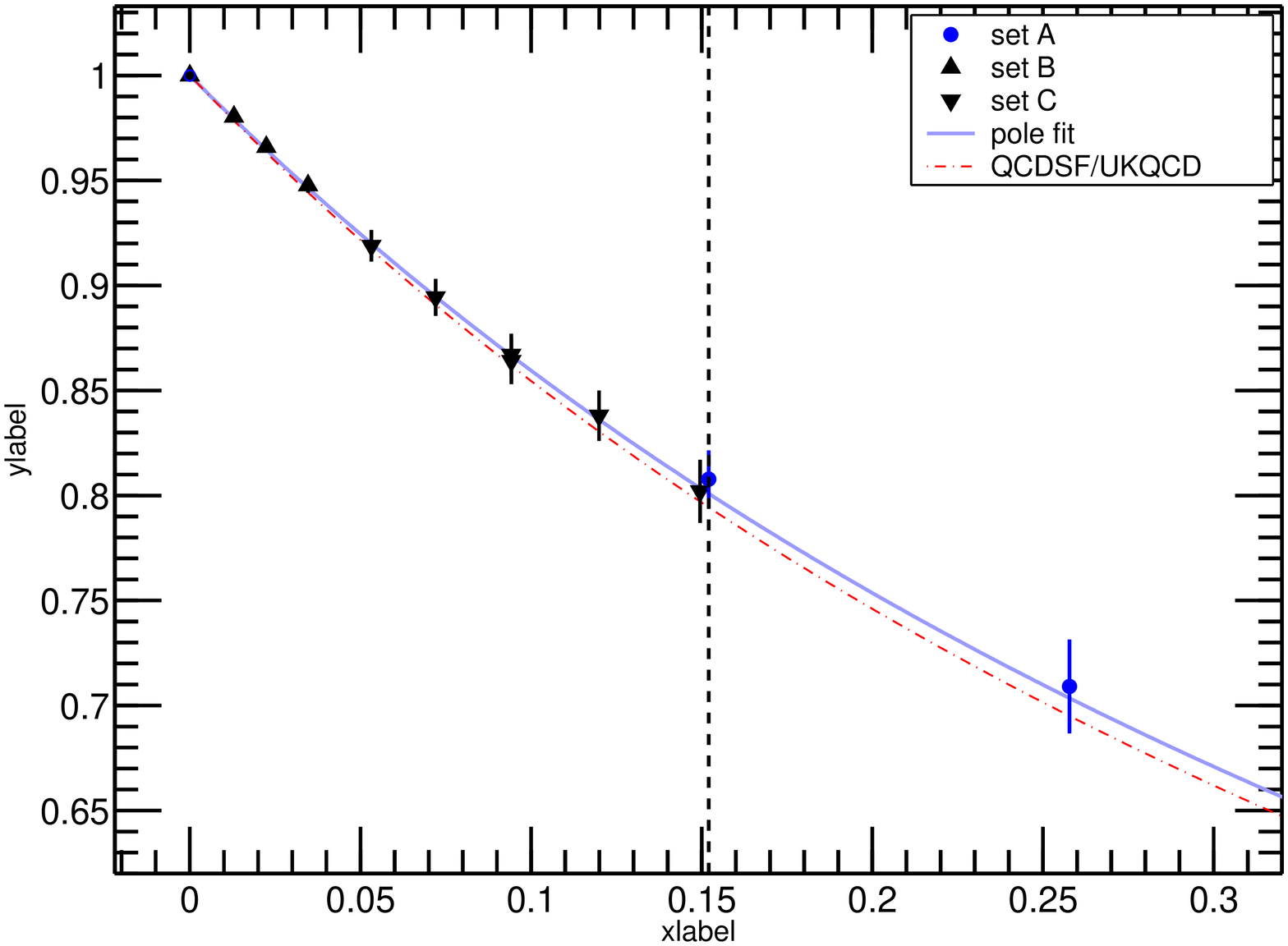}\\[2mm]
\epsfig{scale=.47,angle=-90,file=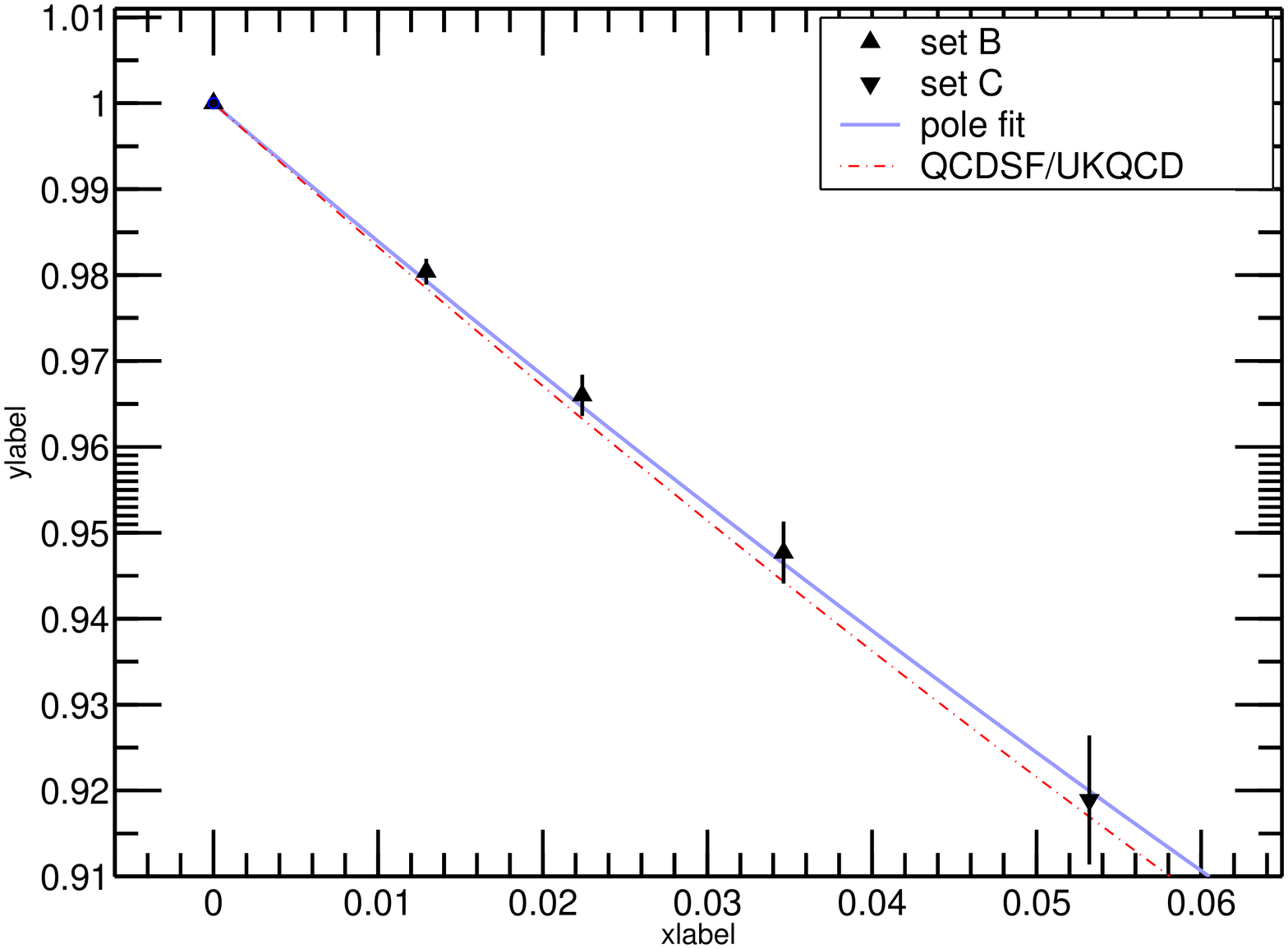}
\end{center}
\caption{Results for the form factor $f^{\pi\pi}(q^2)$ for a pion with
  $m_\pi=330\mev$. The blue dashed curve is a pole fit to the data,
  while the red dashed curve shows the prediction for a $330\mev$ pion
  using results from the QCDSF/UKQCD
  collaboration~\cite{Brommel:2006ww}. The lower plot is a zoom into
  the very low $Q^2$ region.\label{fig:fpipi}}
\end{figure}

The main results of our computation, the form factor of a pion with
$m_\pi=330\mev$ for a range of low values of $Q^2$ (obtained from all
3 data sets A, B and C), are presented in
tab.~\ref{tab:pipi_kinematics} and plotted in fig.~\ref{fig:fpipi}.
The quoted energies have been determined using the continuum
dispersion relation $E(\vec p)=\sqrt{m_\pi^2+\vec p\,^2}$\,, where
$m_\pi$ is the measured pion mass. We also show a zoom of the points
at the lowest values of $Q^2$. The vertical dashed line represents the
position of the lowest value of $Q^2$ which can be reached with
periodic boundary conditions ($Q^2_{\textrm{min}}\simeq 0.15\gev^2$).
From the figure it is satisfying to observe that at
$Q^2_{\textrm{min}}$ the results obtained with partially twisted
boundary boundary conditions join smoothly onto the data obtained by
performing a Fourier sum with momentum of magnitude $2\pi/L$.

Our results from datasets B and C are well represented in the range $0\le Q^2 \le Q^2_{\textrm{min}}$ by the phenomenological pole formula
\begin{equation}\label{eq:poleansatz}
f^{\pi\pi}_{\rm pole}(q^2)=\frac{1}{1-q^2/M_{\rm pole}^2}\,.
\end{equation}
From the slope of the form factor at $q^2=0$ we obtain the pion's electromagnetic charge radius, which is defined by
\begin{equation}
\langle r_\pi^2\rangle \equiv
 6 \frac{d}{d q^2} f^{\pi\pi}(q^2)\big|_{q^2=0}\,.
\end{equation}
The best fit, which is shown as the blue curve in fig.\,\ref{fig:fpipi},
gives $\rpisq_{330\mev} = 6/M_{\rm pole}^2=
0.382(37)(12)(15)\fm^2= \rpisqsimlong\fm^2$, where the first error is
statistical and the second is due to the uncertainty in the lattice
spacing. The third error is to account for our lack of a continuum
extrapolation (as discussed in sec.~\ref{sec:physpiresults} below).
This result corresponds to a pole mass of $(aM_{\rm pole})^2=0.202(20)$.

We compare our results to those of the UKQCD/QCDSF
collaboration~\cite{Brommel:2006ww} who determined the pion form
factor for a number of unphysical pion masses $m_\pi\ge 400\mev$ using periodic boundary conditions. For
each pion mass, they fit their data to the pole form in
(\ref{eq:poleansatz}) and hence determine the dependence of the pole
mass $M_{\rm pole}$ on the pion mass. Their results are well described
by the ansatz,
\begin{equation}
\label{eq:polemass}
M^2(m_{\pi}^2) = c_0 + c_1m^2_{\pi}\,,
\end{equation}
for which they determined $c_0 = 0.517(23)\gev^2$ and $c_1=0.647(30)$.
Thus, for a pion of mass $330\mev$ they predict $\rpisq^{\rm
  UKQCD/QCDSF}_{330\mev} = 0.396(15) \fm^2$. This result, which we
also illustrate in fig.~\ref{fig:fpipi}, is compatible with ours.

\begin{table}[ht]
 \centering
\begin{tabular}[ht]{cccccc}

\hline
\hline data set &maximum $Q^2$ &   linear            &          quadratic           &    cubic      &          pole   \\

\hline  B&0.013 GeV$^2$ & 0.354(28)(11)       &     $-$                      &        $-$    &   0.361(29)(12) \\
	B&0.022 GeV$^2$ & 0.354(26)(11)       &     0.353(35)(11)            &       $-$     &   0.364(27)(12) \\
        B&0.035 GeV$^2$& 0.353(25)(11)       &     0.355(32)(11)            & 0.351(41)(11) &   0.366(27)(12) \\
\hline	C&0.150 GeV$^2$& 0.332(28)(11) 	   &0.387(44)(13) 		&0.406(56)(13) &0.382(37)(12)\\
\hline
\hline
\end{tabular}
\caption{Results for $\rpisq_{\textrm{330\,MeV}}$ obtained by fitting to linear, quadratic or cubic functions of $Q^2$ and by using the pole ansatz (\ref{eq:poleansatz}). In the first row we use only the single point at the lowest value of $Q^2$  ($Q^2=0.013\,$GeV$^2$), in the second we use the two points at the lowest values of $Q^2$ ($Q^2=0.013\,$GeV$^2$ and $Q^2=0.022\,$GeV$^2$) and in the third row we use the points at the lowest three values of $Q^2$ ($Q^2=0.013\,$GeV$^2$, $Q^2=0.022\,$GeV$^2$ and $Q^2=0.035\,$GeV$^2$). The final row corresponds to fits to all 9 points with $Q^2\le Q^2_{\textrm{min}}$. The two quoted errors are statistical and that due to the uncertainty in the lattice spacing.}
\label{tab:poly_fits}
 \end{table}

Although the pole formula (\ref{eq:poleansatz}) is a good representation of our data for the full range $Q^2\le Q^2_{\textrm{min}}$, we find that the points at the smallest values of $Q^2$ tend to give a slightly smaller central
value for the charge radius. We will take as our best 
estimates of $\rpisq_{\textrm{330\,MeV}}$
the value obtained by applying SU(2) ChPT to the points at small $Q^2$ as explained in section~\ref{sec:physical} (we find $\rpisq_{330\,\textrm{MeV}}=\rpisqsu2sim$\,fm$^2$, see tab.\,\ref{tab:resultsSU2}\,). If we limit the fits to the points at small $Q^2$, the slope at $Q^2=0$ (and hence the charge radius) is not sensitive to the precise form of the fitting function. To illustrate this we present in tab.\,\ref{tab:poly_fits} the results obtained by fitting our results for the form factor at the lowest three values of $Q^2$ to the pole form (\ref{eq:poleansatz}) as well as to linear, quadratic and cubic polynomials. In the final row of tab.\,\ref{tab:poly_fits} we present the value of $\rpisq_{\textrm{330\,MeV}}$ obtained by applying the same fits to all 9 points up to $Q^2_{\textrm{min}}$.

\subsection{Comparison of the cost of using point source and
  $\mathbb{Z}(2)$-wall source propagators}
\label{subsec:comparison}

In this study we have used two different formulations of the source in
the computation of the quark propagators. The correlation functions on
data set A have been computed from point source propagators while the
correlation functions on data sets B and C have been computed using
the noise source technique briefly described in
section~\ref{sct:Z2noise}.

In this section we compare the relative computational cost of each
approach in order to achieve similar statistical errors for standard
observables relevant for the phenomenology of light mesons. In a very
similar recent study~\cite{Boyle:2008rh} such a comparison was
carried out for the meson spectrum on a $16^3\times 32$ lattice also
using $N_f=2+1$ Domain Wall fermions and the Iwasaki gauge action. On
this smaller volume the inverse lattice spacing was found to be
$a^{-1}=1.63(3)\gev$ and the study was performed using a pion with
mass $am_\pi \approx 0.44$. The statistical error on the pseudoscalar
and vector meson correlation functions was studied at a fixed
computational cost, i.e. at a fixed number of inversions of the Dirac
matrix. It was found that the stochastic (one-end) approach offers a
factor of two reduction in the error and a definite improvement in
plateau quality over the traditional point source technique.
Preliminary results indicating similar improvements were also reported
by ETMC in \cite{Simula:2007fa}. Here we compare the costs for both
approaches on a larger volume and for a much smaller pion mass. In
particular we perform the comparison for $am_\pi$, $Z_V$ and
$f^{\pi\pi}(q^2=\textrm{--}Q^2_{\textrm{min}})$.
\begin{table}
\begin{center}
\begin{tabular}{ccrrc}
\hline\hline\\[-4mm]
data set & \multicolumn{1}{c}{inversions}&
\multicolumn{1}{c}{$m_\pi$}&
\multicolumn{1}{c}{$Z_V$}&
\multicolumn{1}{c}{$f^{\pi\pi}(\textrm{--}Q^2_{\textrm{min}})$}\\
\hline\\[-4mm]
A&6444&0.1912(7)&0.7148(9)&0.809(14)\\
C&548&0.1910(6)&0.7136(8)&0.802(15)\\
\hline\hline
\end{tabular}
\end{center}
\caption{Comparison of cost and error on quantities relevant for light
	meson phenomenology.}
\label{tab:z2vspoint}
\end{table}
Table~\ref{tab:z2vspoint} shows the results for each quantity for data
sets A and C. In the second column we give the number of inversions of
the Dirac matrix that were carried out in each case. For one
measurement $12$ inversions are necessary in the case of point source
propagators while only one inversion is necessary when using the noise
source technique. On data set A we have $179\times 12$ inversions
times three for the number of sources. Our results for data sets A and
C indicate that the same statistical error for $m_\pi$, $Z_V$ and
$f^{\pi\pi}(\textrm{--}Q^2_{\textrm{min}})$ can be achieved with only about
$1/12$th of the computational cost when using the noise source
technique. (This approximate gain of a factor of $12$ found for this particular simulation should not be confused of course with the 12 inversions performed for each configuration and source for data set A.)

We have also tried to study the error for point-source and
noise-source correlators at fixed cost, i.e. for a given number of
inversions. The cost of the $1176$ measurements which we carried out
with the noise source (data set B) corresponds to $1176/12=98$ point
source measurements. While we could carry out reliable fits to the
correlators on data set B this was not the case for the sub-set of
$98$ measurements of data set A and no quantitative comparison seems
possible. This observation shows however that the statistical
properties of the correlation functions determined with noise-source
propagators are better at the same computational cost.

Very light chiral quarks will display near zero modes associated with topological objects sampled in the ensemble.
Intuitively, we might expect that point
source propagators are more susceptible to the corresponding fluctuations,
particularly if the location of the source is in the vicinity of such
near zero modes. By contrast we expect such outliers to be averaged away
when using a volume source like the one considered in this work. Using
this picture it is not surprising that the gain observed in
tab.~\ref{tab:z2vspoint} goes far beyond the one for the error of the
pion mass observed in~\cite{Boyle:2008rh}. In that work the pion
mass was more than double the one used here, thus the density of near
zero modes was smaller. Furthermore the volume was $(2/3)^3$ of the
one used here, allowing for less volume averaging in the case of the
noise source.

\section{Electromagnetic form factor of a physical pion}
\label{sec:physical}

Having determined the electromagnetic form factor of a pion with
$m_\pi=330\mev$ we can estimate what we would expect for that of a
physical pion. The natural approach to perform this extrapolation is
chiral perturbation theory and in the following subsection we briefly
summarize the predictions of both the $\su2L\times\su2R$ and
$\su3L\times\su3R$ theories.

\subsection{Chiral perturbation theory for the pion electromagnetic form
  factor}

The electromagnetic form factor of the pion has been studied
extensively in both $\su2L\times\su2R$ and $\su3L\times\su3R$ chiral
perturbation theory (ChPT). NLO expressions appear
in~\cite{Gasser:1983yg,Gasser:1984gg} with extensions to NNLO
in~\cite{Bijnens:1998fm,Bijnens:1999hw,Bijnens:2002hp} and we now
briefly summarise the results at NLO. NLO calculations in quenched
ChPT and partially-quenched ChPT appear in~\cite{Arndt:2003ww}. An NLO
calculation with partially twisted boundary conditions in
partially-quenched ChPT exists in~\cite{Jiang:2006gna}; this is
particularly useful to estimate the finite-volume effects. In our
lattice simulation we use unitary points (each valence quark mass is
matched by a sea quark mass) and have small finite-volume effects
(this will be justified below). We therefore use the continuum
(unquenched) QCD results to obtain the form factor of the physical
pion.

Current conservation ensures that $f^{\pi\pi}(0)=1$. At NLO only one
low energy constant (LEC) is relevant for the form factor in both the
$\SU(2)$ and $\SU(3)$ cases. This is denoted by $l_6^r(\mu)$ for
$\SU(2)$ and $L_9^r(\mu)$ for $\SU(3)$ where the superscript $r$
stands for `renormalised' and we have explicitly indicated the
dependence on the renormalization scale $\mu$. The
$\SU(2)$~\cite{Gasser:1983yg} and $\SU(3)$~\cite{Gasser:1984gg}
expressions for the form factor are:
\begin{align}
f^{\pi\pi}_{\SU(2),\mathrm{NLO}}(q^2) &=
 1+\frac1{f^2}\left[
   -2l_6^r \,q^2 + 4\tilde{\mathcal{H}}(m_\pi^2,q^2,\mu^2)\right]
\label{eq:fpipiSU2}\\
f^{\pi\pi}_{\SU(3),\mathrm{NLO}}(q^2) &=
 1+\frac1{f_0^2}\left[
   4L_9^r \,q^2 + 4\tilde{\mathcal{H}}(m_\pi^2,q^2,\mu^2)
             + 2\tilde{\mathcal{H}}(m_K^2,q^2,\mu^2)\right]
\label{eq:fpipiSU3}
\end{align}
where
\begin{equation}
\tilde{\mathcal{H}}(m^2,q^2,\mu^2) =
 \frac{m^2 H(q^2/m^2)}{32\pi^2} -
  \frac{q^2}{192\pi^2}\log\frac{m^2}{\mu^2}
\end{equation}
and
\begin{equation}\label{eq:Hdef}
H(x) \equiv -\frac43 + \frac5{18}x -
 \frac{(x-4)}6 \sqrt{\frac{x-4}x}
 \log\left(\frac{\sqrt{(x-4)/x}\,+1}{\sqrt{(x-4)/x}\,-1}\right)
\end{equation}
with $H(x) = -x/6 + O(x^{3/2})$ for small $x$. For the space-like form
factor considered in this paper $x=q^2/m^2$ is negative and
$(x-4)/x>1$ so that the logarithm in~(\ref{eq:Hdef}) is real as
expected. $f$ and $f_0$ are the pion decay constants in the $\SU(2)$
and $\SU(3)$ chiral limits respectively ($m_u=m_d=0$ with $m_s$ at its
physical value for $\SU(2)$ and $m_u=m_d=m_s=0$ for $\SU(3)$).

The NLO expressions for the charge radius are:
\begin{align}
\langle r_\pi^2\rangle_{\SU(2),\mathrm{NLO}} &=
 -\frac{12l_6^r}{f^2} - \frac1{8\pi^2f^2}
 \Big(\log\frac{m_\pi^2}{\mu^2}+1\Big)\,,\label{eq:rpisqsu2}\\
\langle r_\pi^2\rangle_{\SU(3),\mathrm{NLO}} &=
 \frac{24L_9^r}{f_0^2}
 -\frac1{8\pi^2f_0^2}\Big(\log\frac{m_\pi^2}{\mu^2}+1\Big)
 -\frac1{16\pi^2f_0^2}\Big(\log\frac{m_K^2}{\mu^2}+1\Big)\,.
\label{eq:rpisqsu3}\end{align}
Comparing the expressions for the charge radius gives the relation
between the $\SU(2)$ and $\SU(3)$ NLO LEC's~\cite{Gasser:1984gg}:
\begin{equation}
\label{eq:l6L9relation}
l_6^r(\mu) = -2L_9^r(\mu)
 + \frac1{192\pi^2}\Big(\log\frac{\bar m_K^2}{\mu^2}+1\Big)\,,
\end{equation}
where $\bar m_K^2$ is the kaon mass in the chiral limit for the light
quarks. Using the rho-mass for the renormalization scale,
$\mu=m_\rho$, the second term on the right hand side of this relation
is very small compared to the expected (power-counting) size of the
LECs, so that $l_6^r(m_\rho) \approx -2L_9^r(m_\rho)$. A word of caution should be added however. In deriving eq.\,(\ref{eq:l6L9relation}) from eqs.\,(\ref{eq:rpisqsu2}) and (\ref{eq:rpisqsu3}) we have set $f_0=f$ which is correct at this order. In ref.\,\cite{Allton:2008pn} it was found that $f/f_0\simeq 1.23$ and so we may expect significant corrections to (\ref{eq:l6L9relation}). We follow the approach of ref.\,\cite{Allton:2008pn} and use SU(2)~ChPT to obtain our best results.

The formulae above are obtained in infinite volume. Jiang and Tiburzi
have used partially quenched, partially twisted $\SU(2)$ chiral
perturbation theory to evaluate the finite-volume effects in the case
where only one of the valence quarks is twisted~\cite{Jiang:2006gna}.
This is the case for our three points at the lowest values of $Q^2$
($Q^2=0.013\gev^2$, $Q^2=0.022\gev^2$, and $Q^2=0.035\gev^2$), which
are the points which we use to determine the charge radius and the
LECs $l_6^r$ and $L_9^r$\,. From figures~7 and 8
of~\cite{Jiang:2006gna} we see that for the pion mass
($m_\pi=330\mev$) and volume, $(2.74\fm)^3$, used in our simulation,
the finite volume effects in $\rpisq$ and in $1-f^{\pi\pi}(q^2)$ are
less than $1\%$. Since the remaining errors quoted for these
quantities for a pion with $m_\pi=330\mev$ are $7$--$8\%$, we feel
confident in neglecting the finite volume effects in the remainder of
this analysis. In order to extend the calculations of
ref.~\cite{Jiang:2006gna} to the case in which more than one of the
valence quarks satisfies twisted boundary conditions we would have to
perform SU(3) ChPT calculations in the partially quenched, partially
twisted theory~\cite{Boyle:2007wg}; this is left for a future
publication.

The recent QCDSF/UKQCD two-flavour results for the pion charge
radius~\cite{Brommel:2006ww} include a larger systematic error, a
downwards shift of $6$--$7\%$, arising from finite volume effects. The estimate
of these effects in~\cite{Brommel:2006ww} is obtained using a very different approach to the one we use.
QCDSF/UKQCD fit form factor
data from a range of lattice ensembles, each with a range of pion
masses, to a pole form, eq.~(\ref{eq:poleansatz}), with the pole mass
given by eq.~(\ref{eq:polemass}). A chirally-extrapolated value for
the pole mass translates directly to the chirally-extrapolated result
for the charge radius. Finite volume corrections are modeled by
replacing the expression for the pole mass with
\begin{equation}
M^2(m_{\pi}^2) = c_0 + c_1m^2_\pi + c_2 e^{-m_\pi L},
\end{equation}
where $L$ is the spatial extent of the lattice. For this second form,
additional lattices with varying volumes are added to the fit, but the
results for the lightest pion, $400\mev$, are omitted. A chiral and
infinite volume extrapolation now yields a new physical charge radius,
with the difference quoted as a finite volume systematic error.

We end this subsection with a discussion of another source of uncertainty which the use of chiral perturbation theory can help to estimate. The mass of the (sea) strange quark ($m_s$) in the simulation is different from the physical one ($am_s=0.04$ in the simulation compared to the physical value $0.0343(16)$ found in ref.~\cite{Allton:2008pn}). In SU(3) ChPT we use the mass of the kaon as found from our simulation and hence obtain the value of the LEC $L_9^r$ without the need for further corrections. The LEC $l_6^r$ of SU(2) ChPT on the other hand depends on the mass of the strange quark and, since this is our preferred approach, we need to understand the amount by which $l_6^r$ could be shifted due to the different value of $m_s$. Using eq.\,(\ref{eq:l6L9relation}) and the value of the mass from \cite{Allton:2008pn} to estimate $\bar m_K$, we find that the shift in $l_6^r(m_\rho)$ is about 0.9\% and is hence negligible compared even to the 9\% statistical error (11\% total error) found in section~\ref{sec:physpiresults} below (this is also the case if we use eqs.\,(\ref{eq:rpisqsu2}) and (\ref{eq:rpisqsu3}) without setting $f=f_0$, when the relative error grows to 1.3\%). For the remainder of the analysis we therefore neglect this uncertainty.

\subsection{Results for the physical pion}
\label{sec:physpiresults}

ChPT describes the behaviour of the form factor as a function of both the momentum transfer and the quark masses, providing that these are sufficiently small. We fit our data at fixed quark masses (i.e. for the pion with mass 330\,MeV) as a function of $q^2$ to the NLO formulae for both $\SU(2)$ and $\SU(3)$ ChPT,
eqs.~(\ref{eq:fpipiSU2}) and~(\ref{eq:fpipiSU3}) respectively. In
these fits we use the results $af=0.0665(47)$ and $af_0=0.0541(40)$
which were determined by the RBC/UKQCD collaboration
in~\cite{Allton:2008pn} (in our normalization the decay constant of the physical pion is $f_{\pi^\pm}=130.7(4)\mev$). In this way we obtain the LECs $l_6^r$ and $L_9^r$. Having obtained the LECs in this way, we then use the ChPT formulae given above to determine the form factor (and hence the charge radius) of a physical pion ($m_\pi=139.57\mev$~\cite{Yao:2006px}).

In ref.\cite{Allton:2008pn} it was found that whereas
both $\SU(2)$ and $\SU(3)$ ChPT fit the data for the pion masses and
decay constants, in the $\SU(3)$ case the NLO corrections were very
large, particularly for the decay constant, casting doubt on the convergence of the chiral expansion. For
this reason, in ref.~\cite{Allton:2008pn} the main results were
obtained using $\SU(2)$ ChPT and the above result for the decay
constant in the chiral limit, $af$, includes both the statistical and
systematic errors. The corresponding result for the decay constant in
the $\SU(3)$ limit, $af_0$, on the other hand, includes only the
statistical error.

The results of the chiral extrapolation are summarized in tables
\ref{tab:resultsSU2} and \ref{tab:resultsSU3} for the $\SU(2)$ and
$\SU(3)$ cases respectively. In both tables the first column
corresponds to the result of fitting only to the data point at our
lowest value of $Q^2$ ($Q^2=0.013\gev^2$) to determine the single LEC
($l_6^r(m_\rho)$ or $L_9^r(m_\rho)$) and the charge radius. In the
second column we use the data points at the lowest two values of $Q^2$
($Q^2=0.013\gev^2$ and $0.022\gev^2$) and in the final column we fit the
data for the lowest three values of $Q^2$. The results in the three
columns do not show any dependence on the chosen fit range at these
small values of $Q^2$\,.

Our simulation was performed at a single value of the lattice spacing
and we cannot extrapolate our results to the continuum limit. However,
our action has $O(a^2)$ discretization errors and we
follow~\cite{Allton:2008pn} by assigning a systematic uncertainty of
$4\%$ to measured quantities, representing an estimate of
$(a\Lambda_\mathrm{QCD})^2$ for our lattice spacing. Thus we assign a
$4\%$ error from this source to our values for $1-f^{\pi\pi}(q^2)$.
This relative error is propagated to our results for the LECs and
$\rpisq$, where it appears as the last error quoted in
tables~\ref{tab:resultsSU2} and \ref{tab:resultsSU3}.

\begin{table}
 \[
 \begin{array}{lrrr}
 \hline
 \hline
 Q^2_{\textrm{max}}\,[\gev^2] &
\multicolumn{1}{c}{0.013}&
\multicolumn{1}{c}{0.022}&
\multicolumn{1}{c}{0.035} \\
\hline\\[-4.5mm]
 100\, l_6^r(m_\rho)
 & -0.932(79)(03)(63)(40) & -0.933(73)(03)(63)(40) & -0.932(71)(03)(63)(40) \\
 \rpisq_{330\mev}\,[\fm^2]
 & 0.354(28)(12)(00)(14) & 0.354(26)(12)(00)(14) & 0.354(25)(12)(00)(14) \\
  \rpisq_{\chi}\,[\fm^2]
 & 0.418(28)(12)(04)(14) & 0.419(26)(12)(04)(14) & 0.418(25)(12)(04)(14) \\[1mm]
 \hline
 \hline
 \end{array}
 \]
 \caption{Results from the $\SU(2)$ ChPT fits. The errors are
   statistical, uncertainty in the lattice spacing, uncertainty in
   $af$ and uncertainty from the continuum extrapolation respectively.
   The three columns correspond to using the data at the lowest, the
   lowest two and the lowest three non-zero values of $Q^2$
   respectively, while $Q^2_\textrm{max}$ denotes the largest value of
   $Q^2$ used in the determination.}
 \label{tab:resultsSU2}
\end{table}
\begin{table}
\[
\begin{array}{lrrr}
\hline
\hline
Q^2_{\textrm{max}}\,[\gev^2]   &
\multicolumn{1}{c}{0.013}&
\multicolumn{1}{c}{0.022}&
\multicolumn{1}{c}{0.035} \\
\hline\\[-4.5mm]
100\, L_9^r(m_\rho)
 & 0.307(26)(03)(49)(13) & 0.308(24)(03)(49)(13) & 0.308(23)(03)(49)(13) \\
\rpisq_{330\mev}\,[\fm^2]
 & 0.354(28)(12)(00)(14) & 0.355(26)(12)(00)(14) & 0.355(25)(12)(00)(14) \\
\rpisq_{\chi}\,[\fm^2]
 & 0.460(28)(12)(16)(14) & 0.460(26)(12)(16)(14) & 0.460(25)(12)(16)(14) \\[1mm]
\hline\hline
\end{array}
\]
\caption{Results from the $\SU(3)$ ChPT fits. The errors are
  statistical, uncertainty in the lattice spacing, (statistical)
  uncertainty in $af_0$ and uncertainty from the continuum
  extrapolation respectively. The three columns correspond to using
  the data at the lowest, the lowest two and the lowest three non-zero
  values of $Q^2$ respectively, while $Q^2_\textrm{max}$ denotes the
  largest value of $Q^2$ used in the determination.}
\label{tab:resultsSU3}
\end{table}

Based on the experience of ref.~\cite{Allton:2008pn} and because we
only know the statistical error for $af_0$, we take for our best
estimate the result from the fit to the $\su2L\times\su2R$ expression
at NLO including the three data points at $Q^2=0.013$, $0.022$ and
$0.035\gev^2$,
\begin{equation}
l_6^r(m_\rho)=\lsixr,\qquad \rpisq_{\textrm{330\,MeV}}=0.354(31),\qquad
\rpisq_\chi=\rpisqphyslong\,.
\end{equation}
Comparison of our values for $l_6^r(m_\rho)$ and $L_9^r(m_\rho)$ in
tables~\ref{tab:resultsSU2} and~\ref{tab:resultsSU3} with the
$\SU(2)$--$\SU(3)$ conversion formula in
(\ref{eq:l6L9relation}) reveals deviations up to around $50\%$. By this we mean that the LECs obtained directly from the fit differ from the values extracted
using the conversion formula with the other LEC as input.
Large $\SU(3)$ NLO corrections were seen in the analysis
in~\cite{Allton:2008pn}, and indeed the discrepancy can be reduced very significantly by using eqs.(\ref{eq:rpisqsu2}) and (\ref{eq:rpisqsu3}) without setting $f=f_0$.

In tab.~\ref{tab:other_rsq} we compare our result for the charge
radius to the one determined from experiment and to other recent
computations. Note that the previous lattice results were obtained
with $2$ flavours of sea quarks ($N_f=2$) and using periodic boundary
conditions so that the values of $Q^2$ are much larger
than in this paper.
\begin{table}
 \begin{center}
 \begin{minipage}{.85\linewidth}
 \begin{tabular}{lclllll}
  \hline\hline\\[-4mm]
  collaboration	&&\multicolumn{2}{c}{technique}&
		$\rpisq_\chi [\fm^2]$\\
  \hline\\[-5mm]
  PDG   &\cite{Yao:2006px}&&&0.452(11)\\
  Nam, Kim &\cite{Nam:2007gf}&\multicolumn{2}{l}{instanton vacuum, large $N_c$}
 	&0.455\\
  \hline\\[-4mm]
  QCDSF/UKQCD &\cite{Brommel:2006ww} & $N_f=2$ & Clover & 0.441(19)\\
  JLQCD &\cite{Hashimoto:2005am} & $N_f=2$ & Clover & 0.396(10)\\
  JLQCD &\cite{Kaneko:2007nf} & $N_f=2$ & Overlap & 0.388(15)\\
  RBC/UKQCD & this work & $N_f=2+1$ & Domain Wall & 0.418(31)\\
 \hline\hline
 \end{tabular}
 \end{minipage}
 \begin{minipage}{.128\linewidth}
 \mbox{}\\[-1mm]
 \epsfig{scale=.19,angle=-90,file=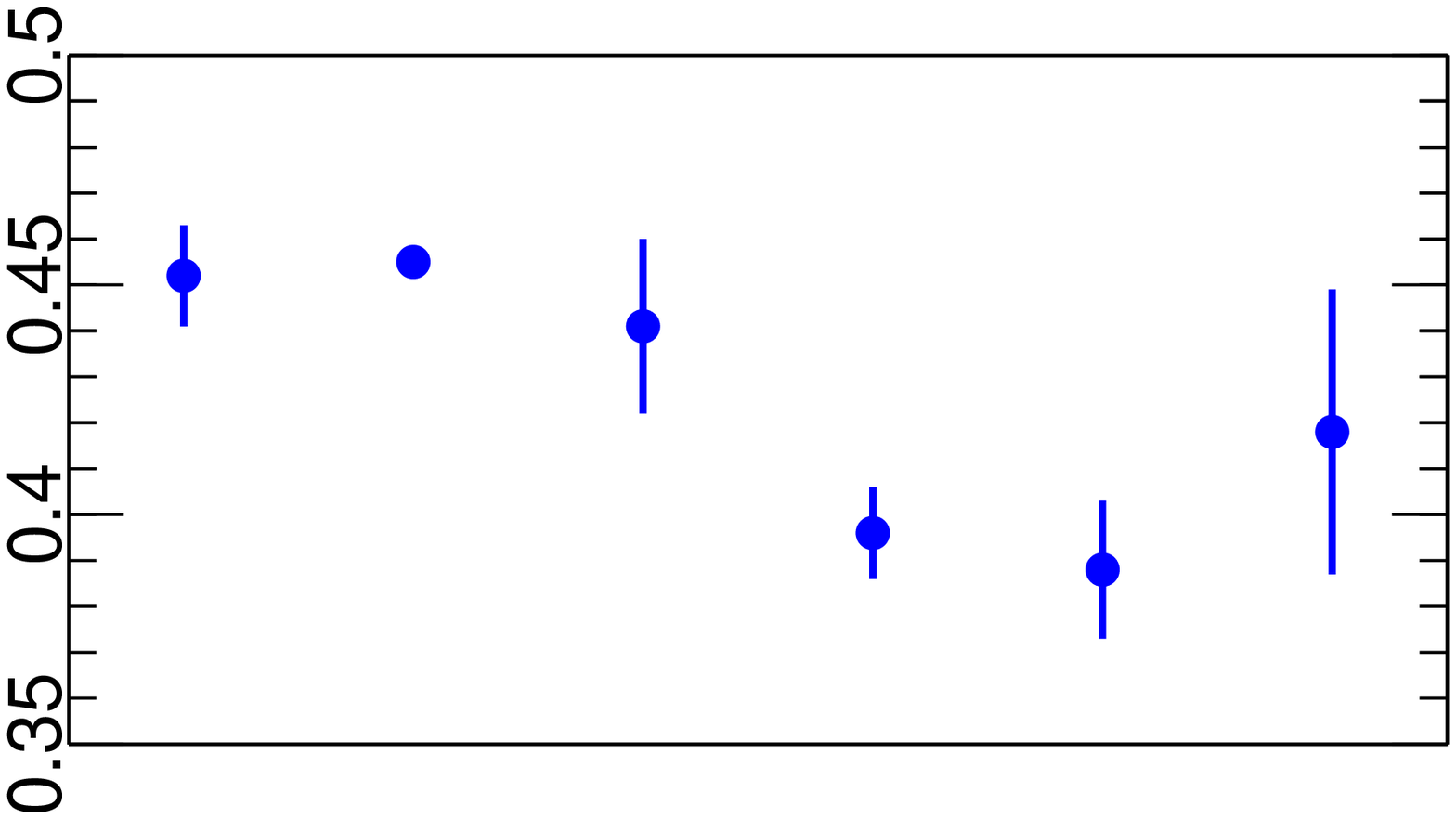}
 \end{minipage}
 \end{center}
 \caption{Previous determinations (excluding quenched lattice results) of the pion's charge radius together
   with the value from the Particle data Group.}
\label{tab:other_rsq}
\end{table}

\begin{figure}
\begin{center}
\psfrag{xlabel}[c][bc][1][0]{$Q^2[\gev^2]$}
\psfrag{ylabel}[c][t][1][0]{$f^{\pi\pi}(q^2)$}
\psfrag{exp}[l][lc][1][0]{\tiny experimental data NA7}
\psfrag{330MeV}[l][lc][1][0]{\tiny lattice data for $m_\pi=330\mev$}
\psfrag{NLO330MeV}[l][lc][1][0]{\tiny $\SU(2)$ NLO lattice-fit; $m_\pi=330\mev$}
\psfrag{NLO139.57MeV}[l][lc][1][0]{\tiny $\SU(2)$ NLO lattice-fit;
		$m_\pi=139.57\mev$}
\psfrag{444OOOOOOOOOOOOOOOOOOOO}[l][lc][1][0]{\tiny $1+\frac 16
		\langle r^2_\pi\rangle^{\rm PDG}Q^2$}
 \epsfig{scale=.4,angle=-90,file=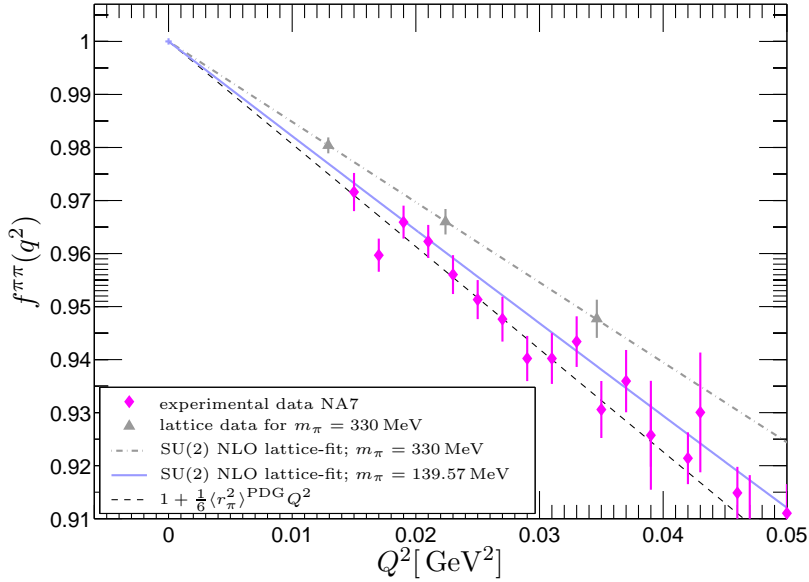}
 \caption{Comparison of experimental results (magenta diamonds) for
   the form factor $f^{\pi\pi}(q^2)$, lattice results at
   $m_\pi=330\mev$ (grey triangles and dash-dotted grey line) and the
   extrapolation of the lattice results to the physical point (blue
   solid line) using NLO $\SU(2)$ chiral perturbation theory. In
   addition we also represent the PDG world average for the charge
   radius using the black dashed line.}
 \label{fig:comparison}
 \end{center}
\end{figure}

In fig.~\ref{fig:comparison} we plot our lattice data for the 330\,MeV pion and the form factor of a physical pion obtained from this data using SU(2)\,ChPT. The experimental data from ref.~\cite{Amendolia:1986wj} is also plotted together with the ChPT formula with the PDG world average for the charge radius (see also tab.\ref{tab:other_rsq}).

\section{Summary and conclusions}\label{sec:concs}

In this paper we have successfully used partially twisted boundary
conditions to compute the electromagnetic form factor of a pion with
mass $330\mev$ at low values of $Q^2$. We use our results to compute the LEC $l_6^r$ of NLO SU(2)
and then to determine the physical form factor and charge
radius, see (\ref{eq:rpisqfinal}). We are able to calculate the
form factor for values of $Q^2$ below the minimum value accessible
with periodic boundary conditions, see fig.~\ref{fig:fpipi}. The
results which we obtain are in good agreement with the experimentally
determined form factor which gives us further confidence in the use of
chiral perturbation theory in the mass range below $330\mev$ (indeed
the value of $f$ which we use in the chiral extrapolation was obtained
with pion masses up to $420\mev$ in ref.~\cite{Allton:2008pn}). The
techniques used in this paper can also be applied to other flavour
non-singlet form factors of mesons and baryons and we strongly
advocate the use of partially twisted boundary conditions in order to
improve significantly the momentum resolution in lattice
phenomenology.

One limitation of the current calculation of the pion's
electromagnetic form factor is that it was performed at a single value
of the lattice spacing, albeit with an action for which the
discretization errors are of $O(a^2)$ and with good chiral and flavour
properties. We are currently generating a set of configurations
with the same action on a $32^3$ lattice with a finer lattice spacing
and will repeat the present calculation with this ensemble. Although the mass and momentum transfers are sufficiently small to expect that NLO SU(2) ChPT is a good approximation, it would be nice to be able to check this explicitly. It is not clear whether in practice a full NNLO calculation can be performed with sufficient precision (i.e. whether the NNLO LECs will be determined sufficiently accurately) but, as it becomes possible to reach lighter quark masses, in the future we will be able to check the stability of the results. The finite-volume corrections for our mass and volume are small~\cite{Jiang:2006gna} and with our precision can be neglected.

In our calculation, we confirm the significant reduction in
computational cost when computing three-point correlation functions
using propagators computed from a single time-slice stochastic source
compared to using point-source propagators.

\acknowledgments
We warmly thank Dirk Br\"ommel for informative discussions about the
content of~\cite{Brommel:2006ww} and B\'alint Jo\'o for help in
learning to use Cray XT4 systems. We are very grateful to the
Engineering and Physical Sciences Research Council (EPSRC) for a
substantial allocation of time on HECToR under the Early User
initiative. We thank Arthur Trew, Stephen Booth and other EPCC HECToR
staff for assistance and EPCC for computer time and assistance on
BlueGene/L.

The calculations also made use of QCDOC computers, and we thank the support staff in the ACF at Edinburgh and at BNL. The QCDOC development and the resulting computer equipment used in this calculation were funded by the U.S.
DOE grant DE-FG02-92ER40699, PPARC JIF grant PPA/J/S/1998/00756 and by RIKEN.
This work was supported by PPARC grants PPA/G/O/2002/00465, PP/D000238/1 and PP/C504386/1.
Our calculations made use of the CHROMA\cite{Edwards:2004sx} and BAGEL software packages.

JMF, AJ, HPdL and CTS acknowledge support from STFC Grant PP/D000211/1
and from EU contract MRTN-CT-2006-035482 (Flavianet). PAB, CK, CMM and
JMZ acknowledge support from STFC grant PP/D000238/1.

\bibliographystyle{JHEP}
\bibliography{tw_pipi}

\end{document}